\documentstyle[preprint,aps,psfig]{revtex}

\begin{document}

\draft

\preprint{}

\title{How to Measure $CP$ Violation in Neutrino Oscillation Experiments?}

\author{Hisakazu Minakata$^1$ and Hiroshi Nunokawa$^2$}

\address{ {$^1$}Department of Physics,
Tokyo Metropolitan University \\
Minami-Osawa, Hachioji, Tokyo 192-03, Japan\\
{$^2$}Instituto de F\'{\i}sica Corpuscular - C.S.I.C.\\
Departament de F\'{\i}sica Te\`orica, Universitat de Val\`encia\\
46100 Burjassot, Val\`encia, Spain\\}

\date{June, 1997}

\preprint{
\parbox{5cm}{
TMUP-HEL-9705\\
FTUV/97-31\\
IFIC/97-30\\
hep-ph/9706281\\
}}
\maketitle
\begin{abstract}
We propose a new method for measuring $CP$ violation in neutrino 
oscillation experiments. The idea is to isolate the term due to 
the $CP$-violating phase out of the oscillation probability by taking 
difference between yields of two (or three) detectors at path-lengths 
$L = 250 \left(\frac{E}{1.35 \mbox{GeV}}\right) 
\left(\frac{\Delta m^2}{10^{-2}\mbox{eV}^2}\right)^{-1} \mbox{km}$
and at $L/3$ (and also at $2L/3$ in the case of three detectors). 
We use possible hierarchies in neutrino masses suggested by 
the astrophysical and the cosmological observations to motivate 
the idea and to examine how the method works.

\end{abstract}
\vskip 0.5cm


$CP$ violation in the lepton sector is an unexplored fascinating 
subject in particle physics. If observed, it should shed light on 
the deep relationship between quarks and leptons, the most fundamental 
structure of matter which we know to date. Moreover, it is suggested 
that $CP$ violation in the lepton sector is the one of the key 
ingredients of the mechanism for generating baryon number asymmetry 
in the universe \cite {FY}. 

A viable way of observing $CP$ violation in the lepton sector 
is to utilize the phenomenon of neutrino oscillation. It was pointed 
out in refs. \cite{Barger,Pakvasa} that the difference between oscillation 
probabilities of the neutrinos and its antineutrinos is proportional 
to the leptonic analogue of the Jarlskog factor \cite {Jarlskog}, 
the unique (in three-flavor neutrinos) phase-convention independent 
measure for $CP$ violation. Recently, measuring $CP$ violation in  
long-baseline neutrino oscillation experiments is of considerable 
interest in the literature \cite {Tanimoto,AS,AKS,MN,BGW}. 

However, there exists a potential obstacle in measuring $CP$ violation 
in long-baseline neutrino experiments. It is the problem of the 
contamination due to the matter effect. Since the earth matter is 
not $CP$ symmetric its effect inevitably produces the fake $CP$ violation 
\cite {matter} which acts as contamination to the genuine $CP$-violating 
effect due to the leptonic Kobayashi-Maskawa phase \cite {KM}. 
Even worse, the matter effect 
dominates over the $CP$ phase effect in a certain region of the mixing 
parameters in the $\nu_\mu \rightarrow \nu_e$ experiment 
\cite {Tanimoto,AS,AKS,MN}. 

In this paper, we suggest a novel way of measuring $CP$ violation 
in neutrino oscillation experiments by proposing the 
multiple detector difference method.\footnote
{Preliminary descriptions of the two-detector difference method
were given in ref. \cite {mina1}. We should mention that the idea of
placing multiple detectors in long-baseline experiments is not new.
For example, it appeared in Brookhaven proposal \cite {BNL}. But
their motivation and the basic idea behind the use of the multiple
detectors is entirely different from ours, and they do not
discuss the possibility of measuring $CP$ violation.}
We will show that our method is relatively free from the problem 
of the matter effect contamination in long-baseline neutrino 
oscillation experiments. 

One way of avoiding the problem of ``matter effect pollution'' is 
to look for the oscillation channel in which genuine $CP$ violating 
effect dominates over the matter effect. This idea is examined in 
detail in ref. \cite{MN} within the restriction of the neutrino mass 
hierarchy to that motivated by hot dark matter and the atmospheric 
neutrino anomaly. It is found that under the constraints from the 
terrestrial experiments the unique case where the $CP$ phase effect 
dominates is the $\nu_\mu \rightarrow \nu_e$ channel in the region of 
large-$s_{13}$ and arbitrary-$s_{23}$ 
(the region (B) to be defined later). 
The $\nu_\mu \rightarrow \nu_\tau$ channel is relatively free 
from matter effect contaminations but they are not negligible.   
Unfortunately, the expected $CP$ violating effect is at most $\sim$ 1\% 
in this type of mass hierarchy due to the strong constraints on mixing 
angles from the terrestrial experiments. 

These discussions of the absolute and the relative magnitudes of $CP$ 
violation is based on the 
$\nu-\bar{\nu}$ difference method.\footnote {Related but different 
proposals in this context have been discussed in ref. \cite {AKS}.}
The major experimental problem with this method is the difficulty 
in determining relative normalization of the neutrino and the 
antineutrino beams. If the $CP$ violation to be measured is of the 
order of a few \% it would require to calibrate the flux of 
neutrino beams to the accuracy better than it, which would be 
extremely difficult, if not impossible, experimentally. 

The multiple detector difference method which we shall discuss in 
this paper aims to overcome this problem. Since the absolute flux of 
the neutrino beam is hard to determine to the accuracy better than 
$\sim$ 10\% \cite {nishi} we may have to give up the comparison 
between the two beams, or two different experiments, if we want 
to do measurement with a few \% level accuracy. 
Therefore, we stick to use a single neutrino beam, $\nu_{\mu}$ 
for example, in the experiment. 
Then, how can one measure $CP$ violation to such precision, or 
even to the accuracy of 1\% level? 

In this paper we confine ourselves into the three-flavor mixing 
scheme of neutrinos. To develop the multiple detector difference 
method we work with the mass hierarchy
\begin{equation}
\Delta M^2 \equiv \Delta m^2_{13} \simeq \Delta m^2_{23} \gg \Delta m^2_{12}
\equiv \Delta m^2,
\label{eqn:hierarchy}
\end{equation}
where $\Delta m^2_{ij} \equiv m^2_j-m^2_i$ ($i,j$ = (1,3), (2,3), (1,2)),
motivated by the solar and the atmospheric neutrino observation 
\cite {solar,atmospheric} and the neutrinos as hot dark matter 
in the mixed dark matter cosmology \cite {hcdm}.

We first focus on the case of mass hierarchy motivated by the 
dark-matter and the atmospheric neutrino observation, 
$\Delta M^2 = 5-100$ eV$^2$ and 
$\Delta m^2 = 10^{-3}-10^{-2}$eV$^2$. 
To find a hint on how to isolate the $CP$ violating term we first 
ignore the matter effect and analyze the structure of the neutrino 
oscillation probability in vacuum. With the mass hierarchy the 
oscillation probability in vacuum for the long-baseline experiments 
can be written as 
\begin{equation}
P(\nu_{\beta} \rightarrow \nu_{\alpha}) = A_{\beta\alpha} + B_{\beta\alpha}
(1-\cos\Delta) + C_{\beta\alpha}\sin\Delta
\label{eqn:probab1}
\end{equation}
with 
\[
\Delta \equiv \frac{\Delta m^2 L}{2E},
\]
where $L$ denotes the path-length of the baseline, 
and $A_{\beta\alpha}, B_{\beta\alpha}$ and $C_{\beta\alpha}$ are 
constants which depend on the mixing angles and the $CP$ phase. 
We note that $C_{\beta\alpha} = 2J$, up to the sign, where $J$ 
indicates the Jarlskog factor whose explicit expression will be 
given later. (See eq.(\ref{eqn:Jarlskog}).) 
The rapid oscillation due to large $\Delta M^2$, 
\begin{equation}
\frac{\Delta M^2 L}{4E} = 127\left(\frac{\Delta M^2}{1\mbox{eV}^2}\right)
\left(\frac{L}{100\mbox{km}}\right)\left(\frac{E}{1\mbox{GeV}}\right)^{-1},
\label{eqn:largeDM}
\end{equation}
is averaged out which produces the first term in (\ref{eqn:probab1}). 

We want to isolate the last term, $J$-term, which is the measure for 
$CP$ violation from the others. It is a simple matter to observe that 
the best way to carry this out is to do the measurements at 
$\Delta = \frac{\pi}{2}$ and $\frac{3}{2}\pi$: 
\begin{eqnarray}
P(\nu_{\beta}\rightarrow\nu_{\alpha};\;\Delta=\frac{3\pi}{2}) &=& 
A_{\beta\alpha} + B_{\beta\alpha} + 2J \nonumber\\
P(\nu_{\beta}\rightarrow\nu_{\alpha};\;\Delta=\frac{\pi}{2}) &=& 
A_{\beta\alpha} + B_{\beta\alpha} - 2J 
\end{eqnarray}
where we took a particular sign for the $J$-term. 
Therefore, the 
difference $\Delta P$ between the oscillation probabilities at 
$\Delta=\frac{3}{2}\pi$ and at 
$\Delta=\frac{\pi}{2}$ 
is nothing but the $CP$ violation $4J$. 

Of course, $\Delta$ is a function of $L/E$. Therefore, the measurement 
of the difference 
$P(\Delta=\frac{3\pi}{2})-P(\Delta=\frac{\pi}{2})\equiv\Delta P_2$ 
can be done either by varying $L$ or $E$, or both.\footnote 
{The possibility of observing $CP$ violation in neutrino oscillation
experiments by measurement at differing path-length, or by varying
energy has been pointed out in ref. \cite {mina2}.}
However, we argue that a measurement of $CP$ violation with accuracy 
better than a few \% would require measurements of $\Delta P_2$ by 
placing two detectors at 
$\Delta =\frac{\pi}{2}$ and $\Delta=\frac{3}{2}\pi$. 
If we try to measure $\Delta P_2$ by varying the energy of the neutrino 
beam with the single detector the energy, of course, have to be varied 
by factor of 3. By such re-adjustment of the neutrino beam energy 
the relative normalization of the beam would become uncertain to the 
order of $\sim$10\%. Therefore, the best thinkable way of avoiding 
the uncertainty of relative normalization of the neutrino flux is 
to do measurement at 2 detectors, using the same neutrino beam, one at 
$\Delta =\frac{3}{2}\pi$, and the other at $\Delta = \frac{\pi}{2}$. 
For the KEK-PS$\rightarrow$Superkamiokande experiment 
in which $L = $250 km, the neutrino beam energy should be tuned to 
$E=1.35 (\frac {\Delta m^2}{10^{-2}eV^2})^{-1}$ GeV 
so that the location of Superkamiokande just corresponds to 
$\Delta=\frac{3}{2}\pi$. 
For the MINOS experiment with $L=730$ km the beam energy to be used 
is $E=3.94 (\frac {\Delta m^2}{10^{-2}eV^2})^{-1}$ GeV. Then, the 
second detector to be build should be located at 1/3 of the baseline, 
$L_2=83.3$ km for the KEK-PS$\rightarrow$Superkamiokande and 
$L_2=243$ km for the MINOS experiments.\footnote{Their geographical 
locations are, respectively, at around the city of Honjo in Saitama 
prefecture and a midpoint between Westfield and Packwaukee, about 
55 Miles north of Madison, Wisconsin.} 

Of course, one cannot make neutrino beam so monochromatic; it must 
have spread in energy. However, it seems possible to make the 
neutrino beam spread as small as $\sim$ 20 \% of the beam energy 
\cite {nishi}. Therefore, at least it is worth to think about the 
possibility. 

Does the matter effect give rise to the serious contamination to 
the genuine $CP$ violating effect in the 2 detector difference method? 
%
%
It appears that the problem has a better shape compared to that in 
the $\nu-\bar\nu$ difference method.  
To understand this point we write down the expression of neutrino 
oscillation probability with the correction of the earth matter 
effect. The expression \cite {MN} is based on the adiabatic 
approximation and is valid to first order in matter perturbation 
theory. If we define the neutrino mixing matrix as 
$\nu_\alpha = U_{\alpha i}\nu_i$ 
the oscillation probability is given by 
\begin{eqnarray}
&&P(\nu_\beta \to \nu_\alpha) = \cr
&& -2 \sum_{i=1,2}\biggl[
\mbox{Re}[U_{\alpha i}U^*_{\alpha3}U^*_{\beta i}U_{\beta3}]
+ \mbox{Re}(UUU\delta V)_{\alpha\beta \: ;\: i3}\biggr]\cr
&& -4 \mbox{Re}[U_{\alpha 1}U^*_{\alpha 2}U^*_{\beta 1}U_{\beta 2}]
\biggl[ \sin^2\biggl(\frac{\Delta m^2}{4E} L\biggr) 
+ \frac{1}{2}aL \biggl(|U_{e2}|^2- |U_{e1}|^2\biggr)
\sin\biggl(\frac{\Delta m^2}{2E} L\biggr) \biggr] \cr
&& -2 J \biggl[\sin\biggl(\frac{\Delta m^2}{2E} L\biggr) 
+ aL \biggl(|U_{e2}|^2- |U_{e1}|^2\biggr)
\cos\biggl(\frac{\Delta m^2}{2E} L\biggr) \biggr]\cr
&& -4 \mbox{Re}(UUU\delta V)_{\alpha\beta \: ;\: 12}
\sin^2\biggl(\frac{\Delta m^2}{4E} L\biggr) \cr
&&  -2 \mbox{Im}(UUU\delta V)_{\alpha\beta \: ;\: 12}
\sin\biggl(\frac{\Delta m^2}{2E} L\biggr)
\label{eqn:probab2}
\end{eqnarray}
where $(UUU\delta V)_{\alpha\beta \: ;\: ij}$ represent first-order 
corrections due to the matter effect and their expressions are given 
in ref. \cite {MN}. More precisely speaking, we made the following 
approximation to derive eq. (\ref {eqn:probab2}): We took average over 
rapid oscillations with the period (\ref {eqn:largeDM}), which 
produces the first two terms in (\ref {eqn:probab2}). We ignored the 
terms of the order of $\frac{Ea}{\Delta M^2}$ because of the 
extreme hierarchy between the dark matter mass scale and the matter 
potential, 
\begin{equation}
\frac{Ea}{\Delta M^2} = 1.04 \times 10^{-4}
\left(\frac{\rho}{2.72\mbox{gcm}^{-3}}\right)
\left(\frac{E}{1\mbox{GeV}}\right) 
\left(\frac{\Delta M^2}{1\mbox{eV}^2}\right)^{-1}.
\end{equation}
We used the constant matter density approximation and ignored the
terms of order $(aL)^2$ or higher, where $L$ is the path length of
the baseline and $a=\sqrt{2}G_FN_e$ with $N_e$ being the electron
number density. We note that
\begin{equation}
aL=0.132\left(\frac{\rho}{2.72\mbox{gcm}^{-3}}\right)
\left(\frac{L}{250\mbox{km}}\right).
\label {eqn:aL}
\end{equation}
Therefore, ignoring $(aL)^2$ term should give a good approximation
at least for the KEK$\to$Superkamiokande experiment.

The difference between probabilities $\Delta P_2$ to be measured at 
2 detectors up to the first order in matter potential $a$ is given by 
\begin{eqnarray}
\Delta P_2(\nu_\beta\to\nu_\alpha) &\equiv& 
P(\nu_{\beta}\rightarrow\nu_{\alpha};\;\Delta=\frac{3}{2}\pi)
-P(\nu_{\beta}\rightarrow\nu_{\alpha};\;\Delta=\frac{\pi}{2})\nonumber\\
&=& 4J
- 8\pi\frac{Ea}{\Delta m^2}
\mbox{Re}[U_{\alpha1}U^*_{\alpha2}U^*_{\beta1}U_{\beta2}]
\cos2\theta_{12} c_{13}^2 \nonumber\\
&& + 8 J \frac{Ea}{\Delta m^2} \cos2\theta_{12} c_{13}^2
\label{eqn:Pdiff}
\end{eqnarray}
where $(\alpha, \beta) = (e, \mu), (\mu, \tau)$ and $(\tau, e)$. 
We have used the standard form of the CKM matrix
\begin{equation}
U=\left[
\begin{array}{ccc}
c_{12}c_{13} & s_{12}c_{13} &   s_{13}e^{-i\delta}\nonumber\\
-s_{12}c_{23}-c_{12}s_{23}s_{13}e^{i\delta} &
c_{12}c_{23}-s_{12}s_{23}s_{13}e^{i\delta} & s_{23}c_{13}\nonumber\\
s_{12}s_{23}-c_{12}c_{23}s_{13}e^{i\delta} &
-c_{12}s_{23}-s_{12}c_{23}s_{13}e^{i\delta} & c_{23}c_{13}\nonumber\\
\end{array}
\right]
\end{equation}
for the neutrino mixing matrix. 
With this parametrization, $J$ is given as 
\begin{equation}
J\equiv \mbox{Im}
[U_{\alpha 1}U_{\alpha 2}^*U_{\beta 1}^*U_{\beta 2}]
=\pm c_{12}s_{12}c_{23}s_{23}c_{13}^2s_{13}\sin\delta,
\label{eqn:Jarlskog}
\end{equation}
where $+$ sign is for cyclic permutations, i.e.,
$(\alpha,\beta)=(e,\mu),(\mu,\tau),(\tau,e)$ and $-$
is for anti-cyclic ones.

We note that in the mass hierarchy with dark matter mass scale the reactor 
and the accelerator experiments constrain the mixing angles into three 
regions on the plane spanned by $s_{13}^2$ and $s_{23}^2$ \cite{raconstr}.
Since our present analysis is motivated by the atmospheric neutrino 
anomaly we restrict ourselves into the two regions (A) or (B) :  
\begin{enumerate}
\renewcommand{\labelenumi}{(\roman{enumi})}
\item[(A)] small-$s_{13}$ and small-$s_{23}$
\item[(B)] large-$s_{13}$ and arbitrary $s_{23}$
\end{enumerate}

In long-baseline experiments, we expect large ($\sim$ order 1) 
oscillation probabilities for $\nu_\mu\rightarrow\nu_e$ and 
$\nu_\mu\rightarrow\nu_\tau$ channels in the regions (A) and (B), 
respectively. For brevity we shall call these channels the 
dominant channels in the respective parameter regions, and the 
alternative ones, i.e., $\nu_\mu\rightarrow\nu_\tau$ in the 
regions (A) etc. the minor channels. Note that 
$-$ 4Re$[U_{\alpha1}U^*_{\alpha2}U^*_{\beta1}U_{\beta2}]$ 
is nothing but the coefficient of 
$\sin^2 (\frac{\Delta m^2}{4E} L)$ term in the oscillation 
probabilities, and therefore is of order unity for the 
dominant channels. 
For $\Delta M^2 = 5$ eV$^2$, $4J$ is at most 
$\sim 10^{-2}$ in both regions (A) and (B). 
See Fig. 1 of ref. \cite {MN}.
We note that $\frac{Ea}{\Delta m^2} \sim 10^{-2}$ in the mass 
hierarchy with which we are dealing. 

Now we can roughly estimate the relative magnitude of the matter and 
the $CP$ violating effects in $\Delta P_2$. It depends upon the regions 
(A) and (B); In the region (A), the $CP$ phase and the matter effects 
are, roughly speaking, comparable unless $\cos2\theta_{12}$ is tuned 
to be small. We have to do the experiments in the minor channel 
$\nu_\mu\rightarrow\nu_\tau$ to avoid matter effect contamination. 
In the region (B) on the other hand, $c_{13}$ is small, 
$c_{13}^2 \sim 10^{-2}$, and therefore the $CP$ phase effect is 
always dominant over matter effect. 
One can do experiments in the dominant channel in the region (B).  
However, to do subtraction between the yields of the end and 
the intermediate detectors it may be better to do experiments 
always in the minor channel; One has to subtract a large number 
from a comparable large number to have a small number in the case 
of dominant channels.  

A special attention has been paid to the long-baseline neutrino 
experiment because with the dark matter motivated mass hierarchy 
the effect of $CP$ violation cannot be observed in short-baseline 
experiments. It can be shown that $CP$ violating effect is suppressed 
by a factor of $\frac {\Delta m^2}{\Delta M^2}$. 

Let us confirm by numerical computation that the qualitative results 
we have obtained so far are correct. 
First we pick up the following two sets of parameters (a) and (b) 
from the allowed regions (A) and (B), respectively, 
\begin{eqnarray}
&\mbox{(a)}& \ \ s_{23}^2  = 3.0\times 10^{-3}, s_{13}^2 = 2.0\times 10^{-2}\\
\label{eqn:seta}
&\mbox{(b)}&\ \ s_{23}^2 = 2.0\times 10^{-2}, s_{13}^2 = 0.98.
\label{eqn:setb}
\end{eqnarray}
They are chosen so that the largest $CP$ violating effect is expected 
in each region (A) or (B), and the same set of parameters are used in 
our analysis done in ref. \cite {MN}. 

In Fig. 1 we present $\Delta P_2(\nu_\mu\to\nu_e)$ 
and $\Delta P_2(\nu_\mu\to\nu_\tau)$ for these 
two parameter sets for the KEK$-$Superkamiokande distance $L=250$ km. 
We have carried out the calculation using the exact solutions obtained 
in ref. \cite{Zaglauer} for a constant matter density and we used 
$\rho$ = 2.72 gcm$^{-3}$ with the electron fraction $Y_e$ = 0.5. 
We took average over the rapid oscillations due to the dark matter 
scale $\Delta M^2$. In the same figure we also plot the values obtained 
by our analytic formula (\ref{eqn:Pdiff}) to indicate that it gives a 
reasonably good approximation.   

In Fig. 2 we present the same quantities but for $L=730$ km, 
i.e., Fermilab-Soudan2 detector distance (MINOS experiment). 
This is important for $\nu_\mu\to\nu_\tau$ channel because $\tau$ 
cannot be produced with such low neutrino energy as $E=1.35$ GeV.
The nice feature of $\Delta P_2$ in Figs. 1 and 2 is that the 
matter effect contamination is relatively smaller than that in the 
$\nu-\bar\nu$ difference method discussed in ref. \cite {MN}.
This is particularly true for $L=250$ km. 
Therefore, if we can measure $\Delta P_2$ to the accuracy of 
$\sim$ 1 \% level it is, in principle, possible to observe $CP$ 
violation in neutrino oscillation experiments.

Now we turn to the question of how the multiple detector difference 
method works for the neutrino mass hierarchy 
motivated by the atmospheric \cite {atmospheric} 
and the solar \cite {solar} neutrino observations, 
$\Delta M^2 = 10^{-3} - 10^{-2} \mbox{eV}^2$ and 
$\Delta m^2 = 10^{-6} - 10^{-4} \mbox{eV}^2$.  
In this case we 
cannot use the matter perturbation theory developed in ref. \cite {MN} 
because $\frac{Ea}{\Delta m^2} \simeq 1-10^2$ cannot be used as an  
expansion parameter. In this paper we rely on the approximate formula 
derived by Arafune, Koike and Sato \cite{AKS} who use as expansion 
parameters $aL$ in (\ref {eqn:aL}) and
\begin{equation}
\frac{\Delta m^2 L}{2E} = 6.4\times 10^{-2} \left(
\frac{\Delta m^2}{10^{-4}\mbox{eV}^2} \right) \left(
\frac{L}{250\mbox{km}} \right) \left(
\frac{E}{1\mbox{GeV}} \right)^{-1}. 
\end{equation}

The general structure of the oscillation probability \cite{AKS} can 
be expressed, to leading order of these expansion parameters, as
\begin{equation}
P(\nu_{\beta}\rightarrow\mu_{\alpha}) 
= \bar{A}_{\beta \alpha}(1-\cos\Delta) 
+ \bar{B}_{\beta \alpha}\Delta\sin\Delta
+ \bar{C}_{\beta \alpha}\Delta(1-\cos\Delta)
\end{equation}
where $\Delta\equiv\frac{\Delta M^2L}{2E}$.
The coefficient of the last term is given by 
$\bar{C}_{\beta \alpha} = 2J\frac{\Delta m^2}{\Delta M^2}$. 
The coefficient $\bar{B}_{\beta \alpha}$ contains the terms 
proportional to either $\frac{\Delta m^2}{\Delta M^2}$ or 
$\frac{Ea}{\Delta M^2}$ and they have similar order of magnitude as 
the last term. Enriched by three terms with different $\Delta$ 
dependence and with similar magnitudes it is unlikely that one can 
separate the last term by using only 2 detectors. Then, we have to 
generalize our method to that of 3 detectors, the ones at 
$\Delta = \pi /2, \pi$, and $3\pi/2$. Then, it is simple to show that 
\begin{equation}
\Delta P_3(\nu_\beta\to\nu_\alpha) \equiv
P(\Delta=\frac{3}{2}\pi) + 3P(\Delta=\frac{\pi}{2})-2P(\Delta=\pi)
= 2\pi J\frac{\Delta m^2}{\Delta M^2}\ .
\label {eqn:Delta-P3}
\end{equation}
To have a feeling of the magnitude of $CP$ violation we choose 
$\Delta m^2 = 10^{-4}\mbox{eV}^2$ and 
$\Delta M^2=10^{-2}\mbox{eV}^2$, and take $s_{12}= 1/2$, 
$s_{23}=1/\sqrt{2}$, and $s_{13}= \sqrt{0.1}$,  
as done by Arafune et al. \cite {AKS}. Then, the RHS of 
(\ref {eqn:Delta-P3}) is estimated to be 0.39$ \times 10^{-2}$.
But, we should note that the parameters we picked up are not those 
which maximize $\Delta P_3$. 

The above discussions on relative and absolute magnitudes of 
$CP$ violations can be confirmed by the precise computation 
of $\Delta P_3$ as we did for $\Delta P_2$. 
We plot in Fig. 3 $\Delta P_3(\nu_\mu\to\nu_e)$ and 
$\Delta P_3(\nu_\mu\to\nu_\tau)$ as a function of $s_{13}$. 
We see that the matter effect contamination is small in 
both channels and $\Delta P_3 \sim 0.5$ for $L=250$ km. 
In Fig. 4 we plot the same as in Fig. 3 but for 
$L=730$ km. In this case, however, the magnitude of matter effect 
can be comparable to the genuine $CP$ effect, which also implies 
that our analytic formula (\ref {eqn:Delta-P3}) is not accurate. 

What are the required statistics for the experiments designed 
for measurements of $CP$ violation at 1 \% level? 
Suppose that we design a long-baseline experiment which will produce 
$N$ muon events at each detector in the absence of oscillation. 
We assume that number of appearance events in the dominant channel 
is $\sim N/2$. Then, the number of events in the minor channel may be  
of the order of $\sim 10^{-3}N$. We have to make subtraction between 
2 or 3 detectors. If we require the accuracy of less than 10 \% 
uncertainty in the number of events in the minor channel it means 
that $N \sim 10^5$. Having such statistics with very narrow-band 
beam is certainly not easy to achieve, but may be possible in 
future Japan Hadron Project and the MINOS experiments.

The potential problem, though technical, with the multiple detector
difference method is that we have to dig a tunnel down from the earth 
surface to build an intermediate detector. We have to dig down to 
1.1 km and 9.3 km for the KEK-PS$\rightarrow$Superkamiokande and the 
MINOS experiments, respectively. We hope that the technical problem can 
be overcome at the stage when the neutrino experiments for measuring 
$CP$ violation is on the time table. 

To establish the multiple detector difference method much more careful 
studies are required on various aspects including beam design. Among 
them one of the most important is the effects of averaging over finite 
energy width of the beam; our preliminary investigation indicates the 
sensitivity of $\Delta P$ against the variation of neutrino energy. 
Keeping these problems in mind, we want to emphasize that this method 
may be the 
only practical way of measuring $CP$ violation to the accuracy of 
a few \% level.  

We thank Koichiro Nishikawa for informative discussions on the 
KEK-PS $\rightarrow$ Superkamiokande experiment. One of us (H.M.) is 
partially supported by Grant-in-Aid for Scientific Research \#09640370 
of the Ministry of Education, Science and Culture, and by Grant-in-Aid 
for Scientific Research \#09045036 under International Scientific 
Research Program, Inter-University Cooperative Research.
The other (H.N.) has been supported by a DGICYT postdoctoral fellowship 
at Universitat de Val\`encia under Grant PB95-1077 
and TMR network ERBFMRXCT960090 of the European Union. 


\newpage
\vglue 0.5cm
\centerline{\hskip 5.0cm
\psfig{file=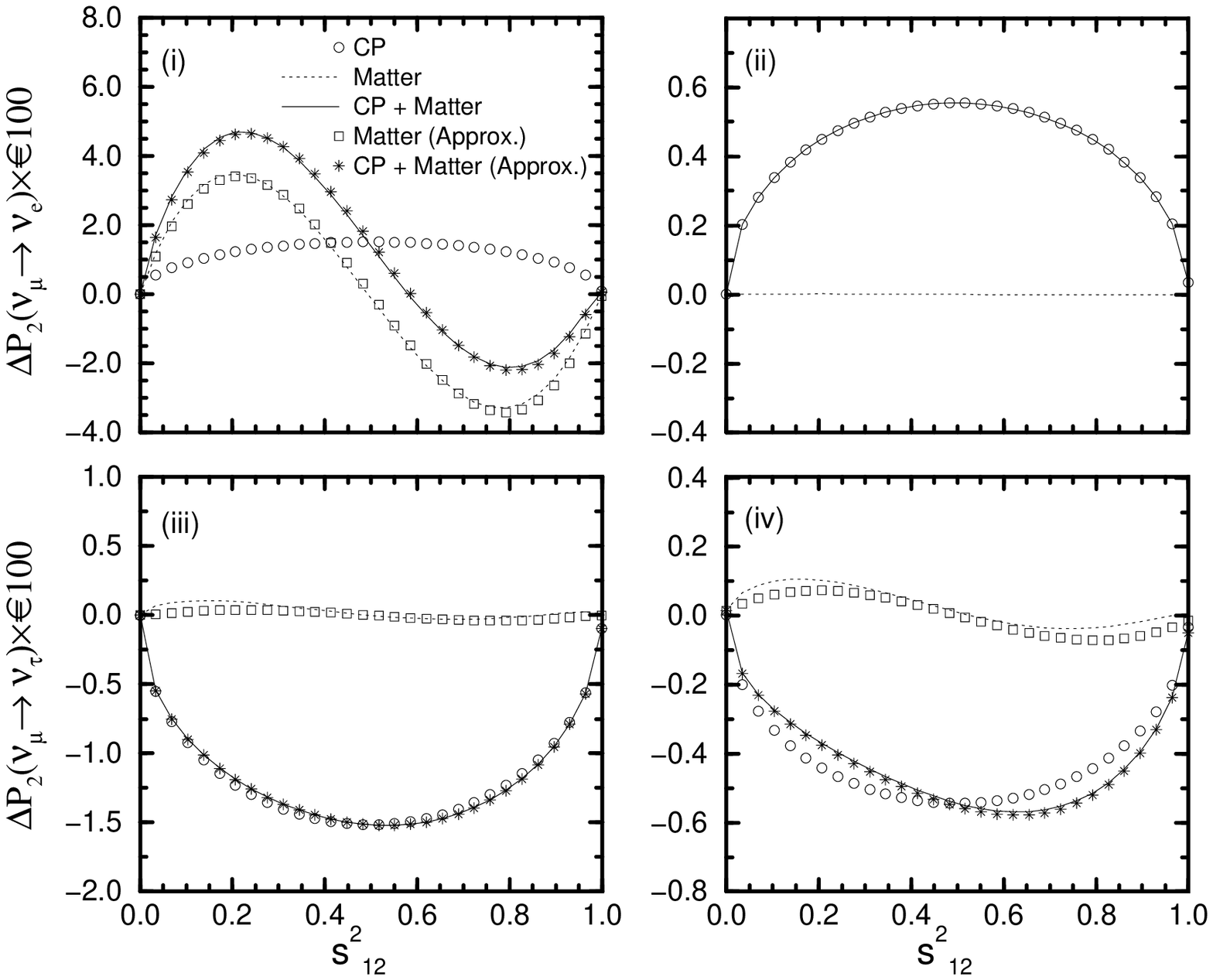,height=18.0cm,width=23.0cm,angle=-90}}
\vglue -5.0cm
\noindent
Fig. 1: We plot in (i) and (ii) $\Delta P_2(\nu_\mu\to\nu_e)$ 
and in (iii) and (iv) $\Delta P_2(\nu_\mu\to\nu_\tau)$ 
as a function of $s_{12}^2$, for the cases where only the genuine 
$CP$ effect (open circles), only the matter effect (dotted lines), 
and both effects (solid lines) exist. We fixed the other mixing 
parameters as $s_{23}^2 = 3.0\times 10^{-3}$, 
$s_{13}^2 = 2.0\times 10^{-2}$ for the left two panels (i) and (iii) 
and $s_{23}^2 = 2.0\times 10^{-2}$, $s_{13}^2$ = 0.98 for 
the right two panels (ii) and (iv). 
The other parameters are fixed to be the same for all the case (i-iv), 
i.e., $E$=1.35 GeV, $\Delta M^2 = 5$ eV$^2$, $\Delta m^2 = 10^{-2}$ eV$^2$ 
and $\delta = \pi/2$ and $L=250$ km. 
We also plot the approximated values for the case where only 
the matter (open squares) and the matter + $CP$ (asterisks) effects 
exist except for (ii) where no appreciable difference between 
the exact and the approximated values can be seen. 
\newpage
\vglue 0.5cm
\centerline{\hskip 5.0cm
\psfig{file=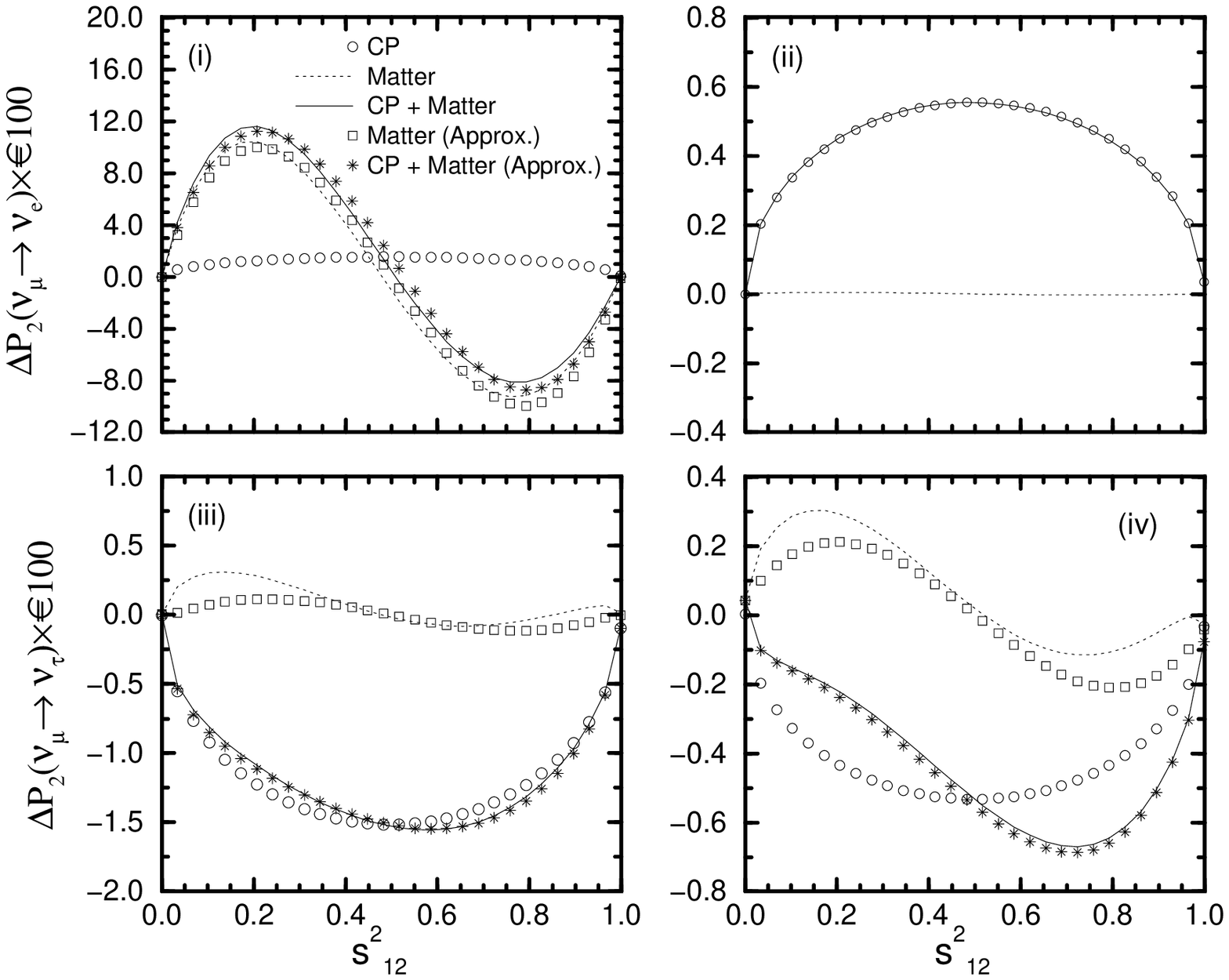,height=18.0cm,width=23.0cm,angle=-90}}
\vglue -5.0cm
\noindent
Fig. 2 : The same as in Fig. 1 but for $L=730$ km and 
$E=3.94$ GeV. 
\newpage
\vglue 0.5cm
\centerline{\hskip 7.0cm
\psfig{file=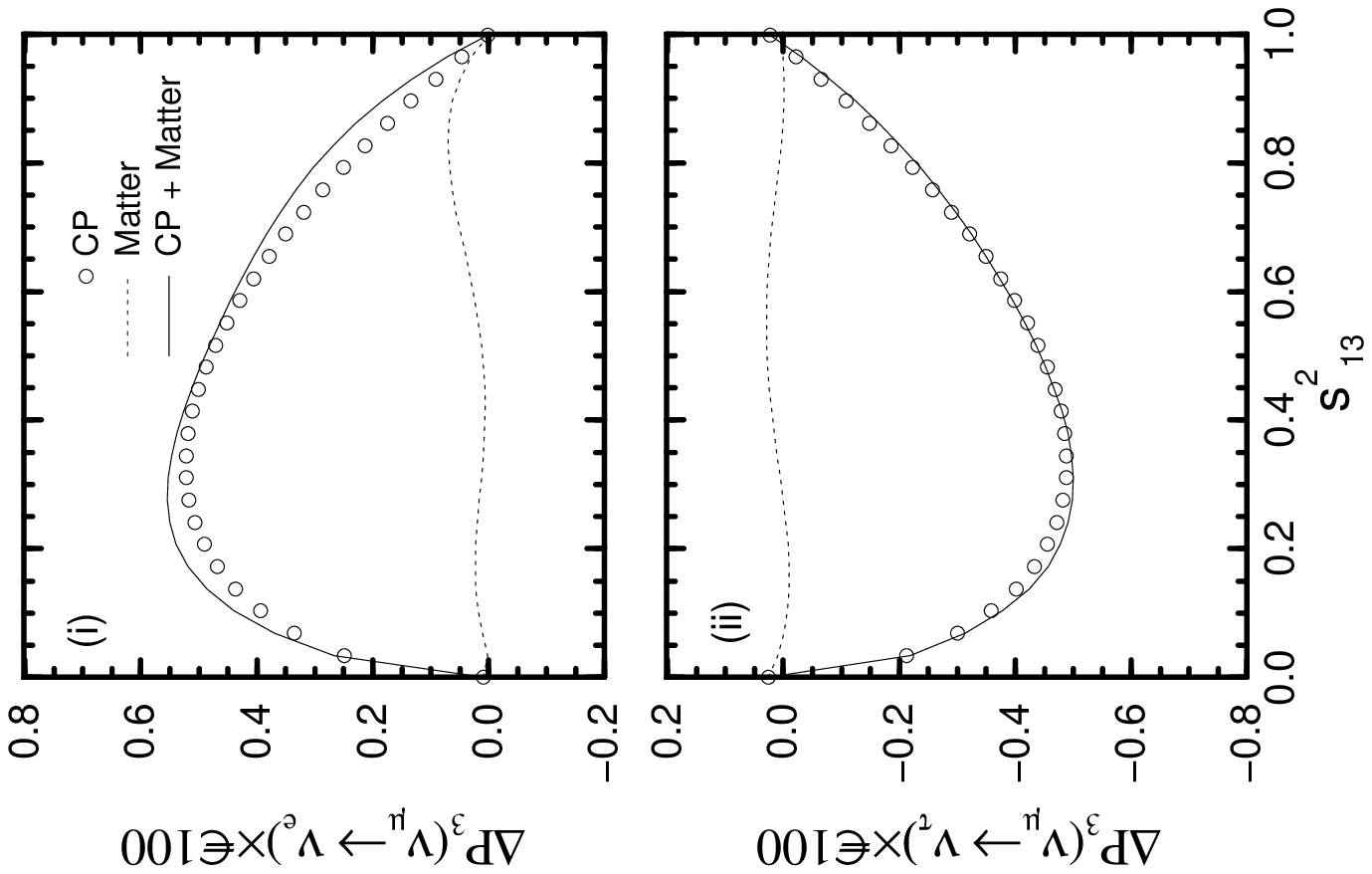,height=18.0cm,width=24.0cm,angle=-90}}
\vglue -5.0cm
\noindent
Fig. 3: We plot in (i) $\Delta P_3(\nu_\mu\to\nu_e)$ 
and in (ii) $\Delta P_3(\nu_\mu\to\nu_\tau)$ as a function of $s_{13}^2$, 
for the cases where only the genuine $CP$ effect (open circles), only 
the matter effect (dotted lines), and both effects (solid lines) exist. 
We fixed the other parameters as $s_{12}^2 = 0.25$, $s_{23}^2 = 0.5$, 
$E$=1.35 GeV, $\Delta M^2 = 10^{-2}$ eV$^2$, $\Delta m^2 = 10^{-4}$ eV$^2$,  
$\delta = \pi/2$ and $L=250$ km. 

\newpage
\vglue 0.5cm
\centerline{\hskip 7.0cm
\psfig{file=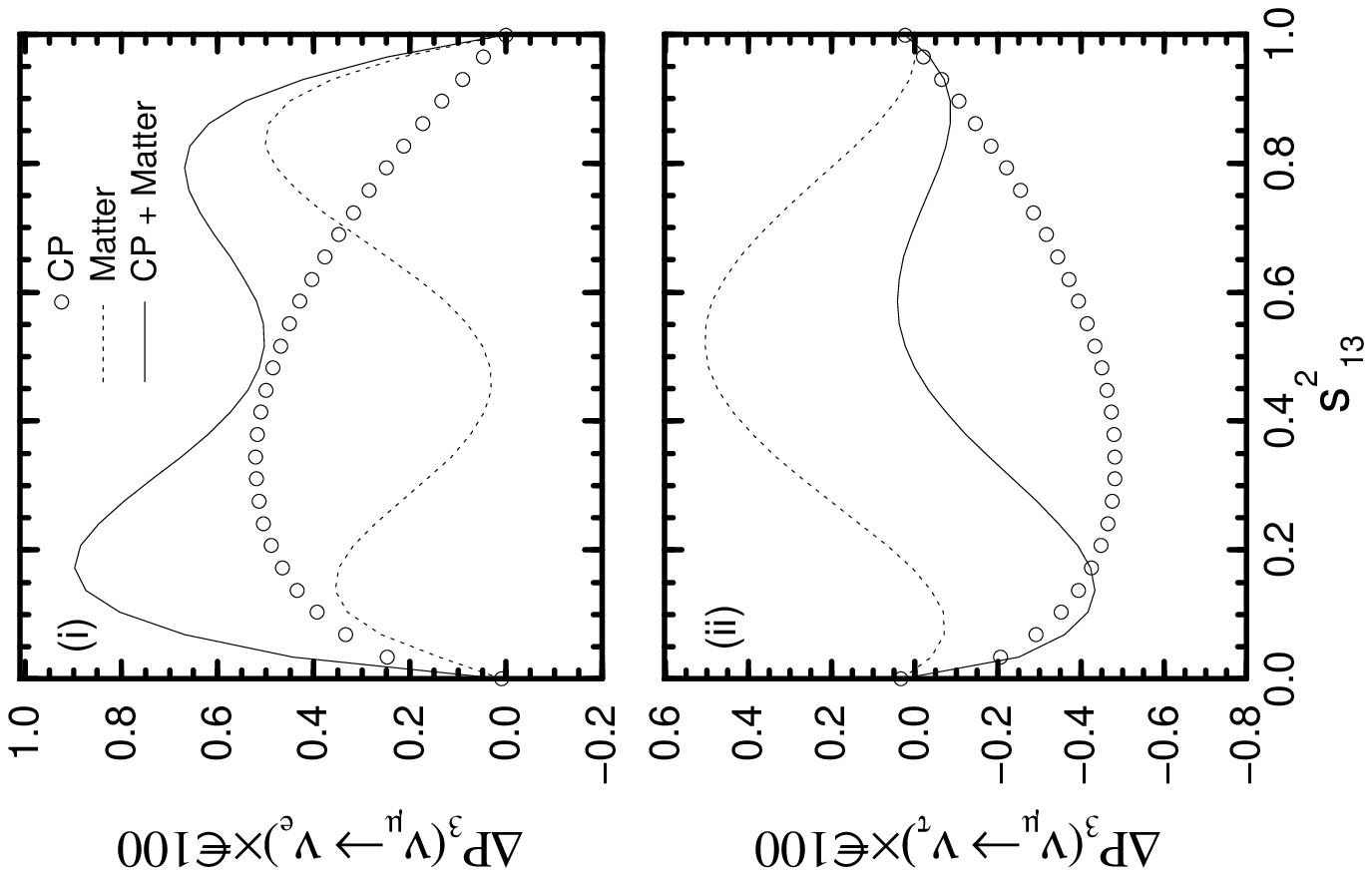,height=18.0cm,width=24.0cm,angle=-90}}
\vglue -5.0cm
\noindent
Fig. 4 : The same as in Fig. 3 but for $L=730$ km and 
$E=3.94$ GeV. 


\begin{thebibliography}{99}

\bibitem {FY}
M. Fukugita and T. Yanagida, Phys. Lett. {\bf B174} (1986) 45.

\bibitem {Barger}
V. Barger, K. Wisnant and R. J. N. Phillips, Phys. Rev. Lett.
{\bf 45} (1980) 2084.

\bibitem {Pakvasa}
S. Pakvasa, in {\it Proceedings of the XXth International Conference
on High Energy Physics}, edited by L. Durand and L. G. Pondrom,
AIP Conf. Proc. No. 68 (AIP, New York, 1981), Vol. 2, pp. 1164.

\bibitem {Jarlskog}
C. Jarlskog, Phys. Rev. Lett. {\bf 55} (1985) 1039.

\bibitem {Tanimoto}
M. Tanimoto, Phys. Rev. {\bf D55} (1997) 322; preprint EHU-96-12, 
hep-ph/9612444.

\bibitem {AS}
J. Arafune and J. Sato, Phys. Rev. {\bf D55} (1997) 1653.

\bibitem {AKS}
J. Arafune, M. Koike and J. Sato, preprint ICRR-385-97-8, 
hep-ph/9703351.

\bibitem {MN}
H. Minakata and H. Nunokawa, preprint TMUP-HEL-9704/FTUV/97-21, 
hep-ph/9705208.

\bibitem {BGW}
S. M. Bilenky, C. Giunti and W. Grimus, preprint UWThPh-1997-11/DFTT 26/97,
hep-ph/9705300.

\bibitem {matter}
T. K. Kuo and J. Pantaleone, Phys. Lett. {\bf B198} (1987) 406;
P. I. Krastev and S. T. Petcov, Phys. Lett. {\bf B205} (1988) 84.

\bibitem {KM}
M. Kobayashi and T. Maskawa, Prog. Theor. Phys. {\bf 49} (1973) 652.

\bibitem {mina1}
H. Minakata, Talk presented at Workshop on Physics of JHP, Yaizu, 
Shizuoka-ken, September 21-23, 1996; 
Talk given at the Conference on Neutrino Physics at Miyako, Miyako, 
Iwate-ken, January 8-10, 1997, 
in {\it Proceedings of the Conference on Neutrino Physics at Miyako}, 
edited by T. Hasegawa, F. Suekane, A. Suzuki, and M. Yamaguchi, 
TOHOKU-HEP-NOTE-97-01, January 1997. 

\bibitem {BNL}
D. Beavis et al. (E889 Collaboration), Physics Design Report, 
BNL No. 52459, April 1995.  

\bibitem {nishi}
K. Nishikawa, private communications. 

\bibitem{solar}
B. T. Cleveland et al., Nucl. Phys. B (Proc. Suppl.) {\bf 38} (1995) 47;
Y. Suzuki, ibid {\bf 38} (1995) 54;
P. Anselmann et al., Phys. Lett. {\bf B285} (1992) 376;
{\bf B314} (1993) 445; {\bf 327}, (1994) 377; {\bf B342}, (1995) 440;
J. N. Abdurashitov et al., Nucl. Phys. B (Proc. Suppl.) {\bf 38} (1995) 60.

\bibitem{atmospheric}
K. S. Hirata et al., Phys. Lett. {\bf B205} (1988) 416;
{\bf B280} (1992) 146;
Y. Fukuda et al., ibid {\bf B335} (1994) 237;
R. Becker-Szendy et al., Phys. Rev. {\bf D46} (1992) 3720;
W. W. M. Allison et al., Phys. Lett. {\bf B391} (1997) 491.

\bibitem {hcdm}
J. A. Holtzman, Astrophys. J. Suppl. {\bf 71} (1989) 1;
J. A. Holtzman and J. R. Primack, Astrophys. J. {\bf 405} (1993) 428;
J. R. Primack, J. Holtzman, A. Klypin, and D. O. Caldwell,
Phys. Rev. Lett. {\bf 74} (1995) 2160;
K. S. Babu, R. K. Schaefer, and Q. Shafi,  Phys. Rev. {\bf D53}
(1996) 606; D. Pogosyan and A. Starobinsky, astro-ph/9502019.


\bibitem {mina2}
H. Minakata, Talk presented at IInd Rencontres du Vietnam, Ho Chi Minh,
Vietnam, October 21-28, 1996, 
in {\it Physics at the Frontiers of the Standard Model}, pp 477, 
edited by Nguyen van Hieu and J. Tran Thanh Van (Editions Frontieres, 
Gif-sur-Yvette, 1996).

\bibitem {raconstr}
H. Minakata, Phys. Rev. {\bf D52} (1995) 6630;
Phys. Lett. {\bf B356} (1995) 61;
S. M. Bilenky, A. Bottino, C. Giunti, and C. W. Kim, 
Phys. Lett. {\bf B356} (1995) 273; 
G. L. Fogli, E. Lisi, and G. Scioscia, Phys. Rev. {\bf D52} (1995) 5334.

\bibitem {Zaglauer}
H. W. Zaglauer and K. H. Schwarzer, Z. Phys. {\bf C40} (1988) 273.

\end{thebibliography}
\end{document}